\documentclass[12pt]{article}
\usepackage{amsfonts,bm}
\usepackage{graphicx}
\baselineskip
\offinterlineskip
\begin{document}

\begin{center}
{\bf An Eigenvalue Problem for a Fermi System and Lie Algebras}
\end{center}

\begin{center}
{\bf  Willi-Hans Steeb$^\dag$ and Yorick Hardy$^\ast$} \\[2ex]

$\dag$
International School for Scientific Computing, \\
University of Johannesburg, Auckland Park 2006, South Africa, \\
e-mail: {\tt steebwilli@gmail.com}\\[2ex]

$\ast$
Department of Mathematical Sciences, \\
University of South Africa, Pretoria, South Africa, \\
e-mail: {\tt hardyy@unisa.ac.za}\\[2ex]
\end{center}

\strut\hfill

{\bf Abstract} We study a Fermi Hamilton operator $\hat K$ 
which does not commute with the number operator $\hat N$.
The eigenvalue problem and the Schr\"odinger equation is solved.
Entanglement is also discussed.
Furthermore the Lie algebra generated by the two terms
of the Hamilton operator is derived and the Lie algebra generated
by the Hamilton operator and the number operator is also classified.

\strut\hfill

\section{Introduction}

In quantum theory Hamilton operators with Fermi-interactions
have a long history \cite{1,2,3,4,5,6,7,8,9}.
Entanglement for Hamilton operators with Fermi-interactions
has been studied by many authors \cite{10,11,12,13}.
Let $c_j^\dagger$, $c_j$ $(j=1,\dots,n)$ be (spin-less) Fermi creation 
and annihilation operators, i.e.
$$
[c_j^\dagger,c_k]_+=\delta_{jk}I, \quad
[c_j,c_k]_+=0,\quad [c_j^\dagger,c_k^\dagger]_+=0 
$$
where $[\,,\,]_+$ denotes the anticommutator and $I$ is the identity
operator. Let $|0\rangle$ be the vacuum state. Then
$c_j|0\rangle=0$ and $\langle 0|0\rangle=1$.
Here we study the self-adjoint Hamilton operator
$$
\hat K = \frac{\hat H}{\hbar\omega} = c_n^\dagger c_{n-1}^\dagger \cdots 
c_2^\dagger c_1^\dagger + c_1c_2 \cdots c_{n-1}c_n\,.
$$
The number operator $\hat N$ is given by
$$
\hat N = \sum_{j=1}^n c_j^\dagger c_j\,.
$$
Obviously $[\hat K,\hat N] \ne 0$. We find the matrix representation of 
$\hat K$ and its eigenvalues and eigenvectors. We utilize the faithful 
matrix representation \cite{6,7,8} for Fermi operators 
\begin{eqnarray*}
c_k^\dagger &=& \overbrace{\sigma_3 \otimes \cdots \otimes \sigma_3 
\otimes \left(\frac12 \sigma_+ \right) \otimes I_2 
\otimes \cdots \otimes I_2 }^{\mbox{$n$-times}} \\ 
c_k &=& \sigma_3 \otimes \cdots \otimes \sigma_3 \otimes
\left(\frac12 \sigma_- \right) \otimes I_2 \otimes \cdots \otimes I_2 \\
&& \qquad \qquad \qquad \qquad k \mbox{-th place} 
\end{eqnarray*}
where $I_2$ is the $2 \times 2$ identity matrix 
and $\sigma_1$, $\sigma_2$, $\sigma_3$ are the Pauli spin matrices
$$
\sigma_1 = \pmatrix { 0 & 1 \cr 1 & 0 }, \quad
\sigma_2 = \pmatrix { 0 & -i \cr i & 0 }, \quad
\sigma_3 = \pmatrix { 1 & 0 \cr 0 & -1 }
$$
with 
$$
\sigma_+ = \sigma_1 + i\sigma_2 = \pmatrix { 0 & 2 \cr 0 & 0 },
\qquad
\sigma_- = \sigma_1 - i\sigma_2 = \pmatrix { 0 & 0 \cr 2 & 0 }\,. 
$$
We also calculate the unitary matrix $U(t)=\exp(-i\hat Ht/\hbar)$ 
to solve the Schr\"odinger and Heisenberg equation of motion. 
Entangled and unentangled states can be found. As entanglement measure 
for the eigenvectors of Hamilton operator $\hat K$ we utilize the entanglement 
measure introduced by Wong and Christensen \cite{14}.
\newline

We also study the Lie algebra generated by
$c_n^\dagger c_{n-1}^\dagger \cdots c_2^\dagger c_1^\dagger$ 
and $c_1c_2 \cdots c_{n-1}c_n$ and the  
Lie algebra generated by the Hamilton operator $\hat K$ and 
the number operator $\hat N$.

\section{Eigenvalue Problem for the Cases $n=1$ and $n=2$}

For the case $n=1$ we have the Hamilton operator
$$
\hat K = \frac{\hat H}{\hbar\omega} = c^\dagger + c = \pmatrix { 0 & 1 \cr 1 & 0 }
= \sigma_1
$$
where the operators $c^\dagger$, $c$ and $c^\dagger c$ are
given by the $2 \times 2$ matrices
$$
c^\dagger = \pmatrix { 0 & 1 \cr 0 & 0 }, \qquad
c = \pmatrix { 0 & 0 \cr 1 & 0 }, \qquad
\hat N = c^\dagger c = \pmatrix { 1 & 0 \cr 0 & 0 } 
$$ 
and the basis is given by
$$
c^\dagger|0\rangle = \pmatrix { 1 \cr 0 }, \qquad 
|0\rangle = \pmatrix { 0 \cr 1 }\,.
$$
Thus the Hamilton operator $\hat H$ acts in the Hilbert space ${\mathbb C}^2$. 
Obviously the eigenvalues of $\hat K$ are $+1$ and $-1$ with the corresponding
normalized eigenvectors (Hadamard basis)
$$
\frac1{\sqrt2} \pmatrix { 1 \cr 1 }, \qquad
\frac1{\sqrt2} \pmatrix { 1 \cr -1 }\,.
$$
For the unitary operator $U(t)=\exp(-i\hat H t/\hbar)$ we obtain
$$
U(t) = \exp(-i\hat H t/\hbar) = 
\pmatrix { \cos(\omega t) & -i\sin(\omega t) \cr 
-i\sin(\omega t) & \cos(\omega t) }\,.
$$
Consider now the case $n=2$. The ordering of the four dimensional basis
is $c_2^\dagger c_1^\dagger|0\rangle$, $c_2^\dagger|0\rangle$,
$c_1^\dagger|0\rangle$, $|0\rangle$.
Utilizing the matrix representation given above we have
$$
c_1 = \frac12 \sigma_- \otimes I_2, \qquad 
c_2 = \sigma_3 \otimes \frac12 \sigma_-\,. 
$$
Thus we obtain the matrix representation
$$
c_1c_2 = \pmatrix { 0 & 0 & 0 & 0 \cr 0 & 0 & 0 & 0 \cr
                    0 & 0 & 0 & 0 \cr 1 & 0 & 0 & 0 }, \quad
c_2^\dagger c_1^\dagger = 
\pmatrix { 0 & 0 & 0 & 1 \cr 0 & 0 & 0 & 0 \cr
           0 & 0 & 0 & 0 \cr 0 & 0 & 0 & 0 }\,.
$$
Consequently
$$
\hat K = c_2^\dagger c_1^\dagger + c_1c_2 =
\pmatrix { 0 & 0 & 0 & 1 \cr 0 & 0 & 0 & 0 \cr
           0 & 0 & 0 & 0 \cr 1 & 0 & 0 & 0 }, \quad
[c_2^\dagger c_1^\dagger,c_1 c_2] = 
\pmatrix { 1 & 0 & 0 & 0 \cr 0 & 0 & 0 & 0 \cr
           0 & 0 & 0 & 0 \cr 0 & 0 & 0 & -1 }\,.
$$
The four eigenvalues of $\hat K$ are $-1$, $1$, $0$ (twice) with the 
corresponding bases for the eigenspaces
$$
\left\{\,\frac1{\sqrt2} \pmatrix { 1 \cr 0 \cr 0 \cr -1 }\,\right\}, \quad
\left\{\,\frac1{\sqrt2} \pmatrix { 1 \cr 0 \cr 0 \cr 1 }\,\right\}, \quad
\left\{\,\pmatrix { 0 \cr 1 \cr 0 \cr 0 }, \,
         \pmatrix { 0 \cr 0 \cr 1 \cr 0 }\,\right\}\,.
$$
The first two eigenspaces have Bell states as basis and are fully entangled 
(except for the zero vector). The last eigenspace consists of unentangled and 
entangled vectors. For the number operator $\hat N$ we find
$$
\hat N = c_1^\dagger c_1 + c_2^\dagger c_2 =
\pmatrix { 2 & 0 & 0 & 0 \cr 0 & 1 & 0 & 0 \cr 0 & 0 & 1 & 0 \cr
0 & 0 & 0 & 0 }
$$ 
with eigenvalues 2, 1 (twice) and 0 and 
$$
\left\{\,\pmatrix{1\cr0\cr0\cr0}\,\right\},\qquad
\left\{\,\pmatrix{0\cr1\cr0\cr0},\,\pmatrix{0\cr0\cr1\cr0}\,\right\},\qquad
\left\{\,\pmatrix{0\cr0\cr0\cr1}\,\right\}
$$
as basis for the respective eigenspaces.
For the unitary operator $U(t)=\exp(-i\hat H t/\hbar)$ we obtain
$$
U(t) = \exp(-i\hat Ht/\hbar) = 
\pmatrix { \cos(\omega t) & 0 & 0 & -i\sin(\omega t) \cr
           0 & 1 & 0 & 0 \cr 0 & 0 & 1 & 0 \cr
          -i\sin(\omega t) & 0 & 0 & \cos(\omega t) }\,.
$$

\section{General Case}

For arbitrary $n$ and $n \ge 2$ the Hamilton operator $\hat K$ is given
by the $2^n \times 2^n$ symmetric matrix over $\mathbb R$
with 1 at the entries $(1,2^n)$ and $(2^n,1)$ and otherwise 0, i.e.
$$
\hat K = 
\pmatrix { 0 & 1 \cr 0 & 0 } \otimes \cdots \otimes \pmatrix { 0 & 1 \cr 0 & 0 } \\ 
+ \pmatrix { 0 & 0 \cr 1 & 0 } \otimes \cdots \otimes  
\pmatrix { 0 & 0 \cr 1 & 0 }\,. 
$$
The commutator $[c_n^\dagger c_{n-1}^\dagger \cdots 
c_2^\dagger c_1^\dagger,c_1c_2 \cdots c_{n-1}c_n]$ admits the matrix
representation
$$
\pmatrix { 1 & 0 \cr 0 & 0 } \otimes \cdots \otimes \pmatrix { 1 & 0 \cr 0 & 0 }  
- \pmatrix { 0 & 0 \cr 0 & 1 } \otimes \cdots \otimes \pmatrix { 0 & 0 \cr 0 & 1 }\,.
$$
This means we have a $2^n \times 2^n$ diagonal matrix
with 1 at the entry $(1,1)$ and $-1$ at the entry $(2^n,2^n)$ and 
otherwise 0. The eigenvalues of $\hat K$ are given by 
$1$, $-1$ and $0$ ($2^n-2$ times). The corresponding bases for the
eigenspaces are
$$
\left\{\,\frac1{\sqrt2} \pmatrix { 1 \cr 0 \cr \vdots \cr 0 \cr 1 }\,\right\}, \quad
\left\{\,\frac1{\sqrt2} \pmatrix { 1 \cr 0 \cr \vdots \cr 0 \cr -1 }\,\right\}, \quad
\left\{\,\pmatrix { 0 \cr 1 \cr \vdots \cr 0 \cr 0 },\, \ldots,\, 
         \pmatrix { 0 \cr 0 \cr \vdots \cr 1 \cr 0 }\,\right\}\,.
$$
The first two eigenspaces consist of entangled vectors (except for the zero vector).
The other $2^n-2$ dimensional eigenspace includes entangled and unentangled vectors.
For the unitary operator $U(t)=\exp(-i\hat H t/\hbar)$ we obtain
$$
\exp(-i\hat Ht/\hbar) = 
\pmatrix { \cos(\omega t) & 0 & \dots & 0 & -i\sin(\omega t) \cr
           0 & 1 & \dots & 0 & 0 \cr 
          \vdots & \vdots & \ddots & \vdots & \vdots \cr
           0 & 0 & \dots & 1 & 0 \cr
          -i\sin(\omega t) & 0 & \dots & 0 & \cos(\omega t) }\,.
$$

\section{Lie Algebras}

We are looking first at the Lie algebra generated by the two operators 
$$
c_n^\dagger c_{n-1}^\dagger \cdots 
c_2^\dagger c_1^\dagger, \qquad c_1c_2 \cdots c_{n-1}c_n\,.
$$
Consider first the case $n=1$. Since $[c^\dagger,c]=2c^\dagger c-I$ and 
$$
[c^\dagger,2c^\dagger c-I] = -2c^\dagger, \quad
[c,2c^\dagger c-I] = 2c 
$$
we find a three-dimensional simple Lie algebra with the basis
$$
c^\dagger, \quad c, \quad c^\dagger c - \frac{I}2\,.
$$
Thus we have a basis of the simple Lie algebra $s\ell(2,{\mathbb R})$.
The matrix representation is
$$
c^\dagger = \pmatrix { 0 & 1 \cr 0 & 0 }, \quad
c = \pmatrix { 0 & 0 \cr 1 & 0 }, \quad 
c^\dagger c-\frac{I}2 = \frac12\pmatrix { 1 & 0 \cr 0 & -1 }\,.
$$
Consider now the case with $n=2$ and the Lie algebra generated by
$c_2^\dagger c_1^\dagger$ and $c_1c_2$. We have
$$
[c_2^\dagger c_1^\dagger,c_1c_2] = c_1^\dagger c_1 +
c_2^\dagger c_2  - I
$$
Next we obtain the commutators 
$$
[c_2^\dagger c_1^\dagger,c_1^\dagger c_1+c_2^\dagger c_2 - I] 
= 2c_1^\dagger c_2^\dagger, 
\qquad
[c_1c_2,c_1^\dagger c_1+c_2^\dagger c_2 - I] = 2c_1c_2\,.
$$
Thus we have a simple three-dimensional Lie algebra with the 
basis $c_2^\dagger c_1^\dagger$, $c_1c_2$, 
$c_1^\dagger c_1+c_2^\dagger c_2 -I$. The Lie algebra is
isomorphic to $s\ell(2,{\mathbb R})$.
The matrix representation given by the diagonal matrix
$$
c_1^\dagger c_1 + c_2^\dagger c_2 - I = 
\pmatrix { 1 & 0 & 0 & 0 \cr 0 & 0 & 0 & 0 \cr
           0 & 0 & 0 & 0 \cr 0 & 0 & 0 & -1 }
$$
and
$$
c_1c_2 = \pmatrix { 0 & 0 & 0 & 0 \cr 0 & 0 & 0 & 0 \cr
                    0 & 0 & 0 & 0 \cr 1 & 0 & 0 & 0 }, \quad
c_2^\dagger c_1^\dagger = 
\pmatrix { 0 & 0 & 0 & 1 \cr 0 & 0 & 0 & 0 \cr
           0 & 0 & 0 & 0 \cr 0 & 0 & 0 & 0 }\,.
$$
Consider now the case $n=3$ and the Lie algebra generated by the
operators $c_3^\dagger c_2^\dagger c_1^\dagger$ and $c_1c_2c_3$. 
We obtain the commutator
$$
[c_3^\dagger c_2^\dagger c_1^\dagger,c_1 c_2 c_3] =
2c_3^\dagger c_2^\dagger c_1^\dagger c_1c_2c_3 
- c_2^\dagger c_1^\dagger c_1 c_2 - c_3^\dagger c_2^\dagger c_2 c_3
- c_3^\dagger c_1^\dagger c_1 c_3 + c_1^\dagger c_1 + c_2^\dagger c_2
+ c_3^\dagger c_3 - I\,.
$$
Next we find
\begin{eqnarray*}
[c_1 c_2 c_3,[c_3^\dagger c_2^\dagger c_1^\dagger,c_1 c_2 c_3]] &=&
2c_1c_2c_3 \cr
[c_3^\dagger c_2^\dagger c_1^\dagger,[c_3^\dagger c_2^\dagger c_1^\dagger,c_1 c_2 c_3]] &=&
-2c_3^\dagger c_2^\dagger c_1^\dagger\,.
\end{eqnarray*}
Thus the three operators 
$c_1c_2c_3$, $c_3^\dagger c_2^\dagger c_1^\dagger$,
$[c_3^\dagger c_2^\dagger c_1^\dagger,c_1 c_2 c_3]$
form a basis of a three-dimensional Lie algebra which is
isomorphic to $s\ell(2,{\mathbb R})$.
The matrix representation of 
$[c_3^\dagger c_2^\dagger c_1^\dagger,c_1c_2c_3]$ is given by
$$ 
\pmatrix { 1 & 0 \cr 0 & 0 } \otimes \pmatrix { 1 & 0 \cr 0 & 0 }
\otimes \pmatrix { 1 & 0 \cr 0 & 0 } 
- \pmatrix { 0 & 0 \cr 0 & 1 } \otimes \pmatrix { 0 & 0 \cr 0 & 1 }
\otimes \pmatrix { 0 & 0 \cr 0 & 1 }\,.  
$$
For arbitrary $n$ we have
\begin{eqnarray*}
[c_1 \cdots c_n,[c_n^\dagger \cdots c_1^\dagger,c_1 \cdots c_n]] &=&
2c_1 \cdots c_n \cr
[c_n^\dagger \cdots c_1^\dagger,[c_n^\dagger \cdots c_1^\dagger,
c_1 \cdots c_n]] &=& -2c_n^\dagger \cdots c_1^\dagger\,.
\end{eqnarray*}
Thus for arbitrary $n$ the three operators 
$$
c_n^\dagger c_{n-1}^\dagger \cdots 
c_2^\dagger c_1^\dagger, \qquad c_1c_2 \cdots c_{n-1}c_n, \qquad
[c_n^\dagger c_{n-1}^\dagger \cdots 
c_2^\dagger c_1^\dagger,c_1c_2 \cdots c_{n-1}c_n]
$$
form a basis of a three dimensional Lie algebra which is isomorphic
to $s\ell(2,{\mathbb R})$.
\newline

Next we study the Lie algebra generated by the Hamilton operator
$\hat K$ and the number operator $\hat N$. Let $n=1$. We find for the 
commutators
$$
[\hat K,\hat N] = c - c^\dagger, \quad 
[\hat K,c-c^\dagger] = 4c^\dagger c - 2I = 4\hat N - 2I, \quad 
[\hat N,c-c^\dagger] = -c^\dagger - c = -\hat K\,. 
$$
Thus we have a four-dimensional non-commutative Lie algebra
with a basis given by $\hat K$, $\hat N$, $c-c^\dagger$, $I$.
Owing to the operator $I$ the Lie algebra is not semisimple.
Utilizing the matrix representation we have
$$
\hat K = \pmatrix { 0 & 1 \cr 1 & 0 } = \sigma_1, \quad
\hat N = \pmatrix { 1 & 0 \cr 0 & 0 }, \quad
c - c^\dagger = \pmatrix { 0 & -1 \cr 1 & 0 } = -i\sigma_1, \quad
I = \pmatrix { 1 & 0 \cr 0 & 1 }\,.
$$
For $n=2$ we have the commutators
\begin{eqnarray*}
[\hat K,\hat N] &=& 2(c_1c_2-c_2^\dagger c_1^\dagger) \cr
[\hat K,[\hat K,\hat N]] &=& 4(\hat N-I) \cr
[\hat N,[\hat K,\hat N]] &=& -4\hat K \cr
[\hat K, [\hat K,[\hat K,\hat N]]] &=& 4[\hat K, \hat N]\cr
[\hat N, [\hat K,[\hat K,\hat N]]] &=& 0\cr
[[\hat K, \hat N], [\hat K,[\hat K,\hat N]]] &=& 16 \hat K\,.
\end{eqnarray*}
Thus the operators $\hat K$, $\hat N$, $c_1c_2-c_2^\dagger c_1^\dagger$,
$I$ form a basis of the four dimensional Lie algebra which is
not semisimple owing to $I$. The matrix representation is
$$
\hat K = \pmatrix { 0 & 0 & 0 & 1 \cr 0 & 0 & 0 & 0 \cr
                    0 & 0 & 0 & 0 \cr 1 & 0 & 0 & 0 }, \quad
\hat N = \pmatrix { 2 & 0 & 0 & 0 \cr 0 & 1 & 0 & 0 \cr
                    0 & 0 & 1 & 0 \cr 0 & 0 & 0 & 0 }, \quad
c_1c_2-c_2^\dagger c_1^\dagger = 
\pmatrix { 0 & 0 & 0 & -1 \cr 0 & 0 & 0 & 0 \cr 0 & 0 & 0 & 0 \cr
           1 & 0 & 0 & 0 }\,.
$$ 
For $n=3$ we have the commutators
\begin{eqnarray*}
[\hat K,\hat N] &=& 3(c_1c_2c_3 - c_3^\dagger c_2^\dagger c_1^\dagger) \cr
[\hat K,[\hat K,\hat N]] &=& 
6[c_3^\dagger c_2^\dagger c_1^\dagger,c_1 c_2 c_3] \cr
[\hat N,[\hat K,\hat N]] &=& -9\hat K \cr
[\hat K, [\hat K,[\hat K,\hat N]]] &=& 4[\hat K, \hat N]\cr
[\hat N, [\hat K,[\hat K,\hat N]]] &=& 0\cr
[[\hat K, \hat N], [\hat K,[\hat K,\hat N]]] &=& 36 \hat K\,.
\end{eqnarray*}
Thus we have a four dimensional Lie algebra given by the operators
$\hat K$, $\hat N$, $[\hat K,\hat N]$, $[\hat K,[\hat K,\hat N]]$.
The Lie algebra is not semisimple.
\newline

For general $n$ we have 
\begin{eqnarray*}
[\hat K,\hat N] &=& n(c_1\cdots c_n-c_n^\dagger \cdots c_1^\dagger) \cr
[\hat K,[\hat K,\hat N]] &=& 2n[c_n^\dagger \cdots c_1^\dagger,
c_1 \cdots c_n] \cr
[\hat N,[\hat K,\hat N]] &=& -n^2 \hat K \cr
[\hat K, [\hat K,[\hat K,\hat N]]] &=& 4[\hat K, \hat N]\cr
[\hat N, [\hat K,[\hat K,\hat N]]] &=& 0\cr
[[\hat K, \hat N], [\hat K,[\hat K,\hat N]]] &=& 4n^2 \hat K\,.
\end{eqnarray*}
Thus we find that the four operators $\hat K$, $\hat N$, 
$[\hat K,\hat N]$, $[\hat K,[\hat K,\hat N]]$ provide a basis 
of a four dimensional Lie algebra which is not semisimple.  

\section{Entanglement}

An $n$-tangle \cite{14,9} can be defined for the finite dimensional
Hilbert space ${\cal H}={\mathbb C}^{2^n}$, with $n=3$ or $n$ even.
Consider the finite-dimensional Hilbert space ${\cal H}={\mathbb C}^{2^n}$ 
and the normalized states 
$$
|\psi\rangle = \sum_{j_1,j_2,\dots,j_n=0}^1 
c_{j_1,j_2,\dots,j_n}|j_1\rangle \otimes |j_2\rangle \otimes \cdots
\otimes |j_n\rangle 
$$
in this Hilbert space. Here $|0\rangle$, $|1\rangle$ denotes the 
standard basis. Let $\epsilon_{jk}$ $(j,k=0,1)$ be defined by 
$\epsilon_{00}=\epsilon_{11}=0$, $\epsilon_{01}=1$, 
$\epsilon_{10}=-1$. Let $n$ be even or $n=3$. Then an 
$n$-tangle can be introduced by
$$
\begin{array}{ccl}
\tau_{1\dots n} &=& 2 \Bigg| 
\displaystyle\!\!\!\sum_{{\scriptstyle \alpha_1,\dots,\alpha_n=0 \atop
\scriptstyle \ldots } \atop \scriptstyle \delta_1,\dots,\delta_n=0}^1
\!\!\!\!\!\!\!\!c_{\alpha_1 \dots \alpha_n} c_{\beta_1 \dots \beta_n} 
c_{\gamma_1 \dots \gamma_n} c_{\delta_1 \dots \delta_n} \Bigg. \\
&& \displaystyle\Bigg. \times \epsilon_{\alpha_1\beta_1} \epsilon_{\alpha_2\beta_2} 
\cdots \epsilon_{\alpha_{n-1}\beta_{n-1}} \epsilon_{\gamma_1\delta_1}
\epsilon_{\gamma_2\delta_2} \cdots \epsilon_{\gamma_{n-1}\delta_{n-1}}
\epsilon_{\alpha_n\gamma_n} \epsilon_{\beta_n\delta_n}\Bigg|.
\end{array}
$$
This includes the definition for the 3-tangle with $n=3$.
\newline

Consider now the eigenvectors of the Hamilton operator $\hat K$
with $n \ge 2$. Then the eigenvectors belonging to $-1$ and $+1$
are fully entangled and include part of the Bell basis. 
The eigenspace belonging to the eigenvalue 0 consists of both
entangled and unentangled vectors.

\section{Conclusion}

We have studied a Fermi Hamilton operator. If the eigenvalues are 
degenerate then by linear combinations we can construct entangled
states from unentangled states. A computer algebra program 
written in SymbolicC++\cite{15} for the manipulation of the Fermi operators
is available from the authors.
\newline

The model described above has a straightforward
extension to the Fermi Hamilton operator with spin
$$
\hat K = \frac{\hat H}{\hbar\omega} = 
c_{n\uparrow}^\dagger c_{n-1\uparrow}^\dagger \cdots 
c_{2\uparrow}^\dagger c_{1\uparrow}^\dagger + 
c_{n\downarrow}^\dagger c_{n-1\downarrow}^\dagger \cdots 
c_{2\downarrow}^\dagger c_{1\downarrow}^\dagger 
+ c_{1\uparrow}c_{2\uparrow} \cdots c_{n-1\uparrow}c_{n\uparrow}
+ c_{1\downarrow}c_{2\downarrow} \cdots c_{n-1\downarrow}c_{n\downarrow}
$$
with the number operator $\hat N$ and spin operator $\hat S_z$ given by
$$
\hat N = \sum_{j=1}^n (c_{j\uparrow}^\dagger c_{j\uparrow}
+c_{j\downarrow}^\dagger c_{j\downarrow}), \qquad
\hat S_z = \frac12 \left(\sum_{j=1}^n (c_{j\uparrow}^\dagger c_{j\uparrow}
-c_{j\downarrow}^\dagger c_{j\downarrow})\right)
$$
where $[\hat K,\hat N] \neq 0$, $[\hat K,\hat S_z] \ne 0$,
$[\hat N,\hat S_z]=0$.
Here the matrix representation is given by \cite{6,7,8,9}
\begin{eqnarray*}
&& \qquad \qquad \qquad k \mbox{-th place}  \\
c_{k\uparrow}^\dagger &=& \underbrace{\sigma_3 \otimes \cdots \otimes 
\sigma_3 \otimes \left(\frac12 \sigma_+\right) 
\otimes I_2 \otimes \cdots \otimes I_2}_{2n\ \mbox{times}} \\ [0.1 cm]
c_{k\downarrow}^\dagger &=& \sigma_3 \otimes \cdots \otimes \sigma_3 \otimes \left(
\frac12 \sigma_+ \right) \otimes I_2 \otimes \cdots \otimes I_2 \\
&& \quad \qquad \qquad \qquad (k+n) \mbox{-th place}
\end{eqnarray*}
where $k=1,\dots,n$. Whereas the Fermi system discussed above provides 
a three energy level system for $n=2$ with eigenvalues $1$, $0$, $-1$
including the spin provides five energy levels with 
eigenvalues $2$, $1$, $0$, $-1$, $-2$ for $n=2$.
\newline

The model discussed above can easily be extended to Majorana fermions
on a lattice \cite{16,17,18,19}. 
Given a set of $n$ (spin-less) Fermi creation and annihilations
operators $c_j^\dagger$, $c_j$ $(j=1,\dots,n)$ we can define
the set of $2n$ (real) Majorana fermion operators on a lattice 
$\gamma_{j1}$, $\gamma_{j2}$ $(j=1,\dots,n)$ as
$$
c_j = \frac12(\gamma_{j1}+i\gamma_{j2}), \qquad
c_j^\dagger = \frac12(\gamma_{j1}-i\gamma_{j2})
$$
where $\gamma_{j1}^*=\gamma_{j1}$, $\gamma_{j2}^*=\gamma_{j2}$
and $\gamma_{j1}^2=\gamma_{j2}^2=I$. It follows that
$$
\gamma_{j1} = c_j^\dagger + c_j, \qquad
\gamma_{j2} = i(c_j^\dagger - c_j)
$$
with 
$$
[\gamma_{j1},\gamma_{j2}]_+=0, \quad 
[\gamma_{j1},\gamma_{j1}]_+=2I, \quad
[\gamma_{j2},\gamma_{j2}]_+=2I
$$
and $[\gamma_{j\ell},\gamma_{k\ell'}]_+=0$ for $j \ne k$.
Another extension of the Hamilton operator would be to consider
$$
\hat H = J(b^\dagger \otimes c_n^\dagger c_{n-1}^\dagger \cdots 
c_2^\dagger c_1^\dagger + b \otimes c_1c_2 \cdots c_{n-1}c_n)
$$
where $b^\dagger$, $b$ are Bose creation and annihilation operators
with the commutation relation $[b,b^\dagger]=I$.
\newline

{\bf Acknowledgment}
\newline

The authors are supported by the National Research Foundation (NRF),
South Africa. This work is based upon research supported by the National
Research Foundation. Any opinion, findings and conclusions or recommendations
expressed in this material are those of the author(s) and therefore the
NRF do not accept any liability in regard thereto.

\end{document}